# Comment on « Molecular dynamics study of the threshold displacement energy in vanadium »

by
P. Vajda
Laboratoire des Solides Irradiés, Ecole Polytechnique, Palaiseau, France

Abstract

The simulation study on Frenkel pair creation in vanadium by L.A. Zepeda-Ruiz et al. (Phys. Rev. B 67, 134114, 2003) is criticized for its lack of reference to existing experimental work and for generally inadequate treatment of literature data. Hence, the validity of the ensuing discussion of various results (e.g. the value of the minimum TDE and the Frenkel pair stability in V) is questioned.

I am much surprised about the above paper [1] as concerns the inadequate way of the literature treatment by its authors and which has apparently not been remarked by the referees. It seems a typical example of how experimental work is ignored by computer simulation teams and poses the general question about the validity of thus obtained data when compared to « real life ». It is not « because of the intrinsic difficulty in obtaining experimental data » (3rd paragraph of the Introduction in [1]) that one has to ignore them. (It is quite significant, for example, that one of the only two cited in the Introduction references on experimental determination of the threshold displacement energy (TDE) in metals – namely ref. [Z-10], by Zinkle and Kinoshita[2] - is a review article on defect production in *ceramics*.) And since « a critical need exists for the prediction of the properties and evolution of vanadium alloys under fusion reactor conditions » (Introduction of [1]), the authors should have checked the already existing literature before stating « no work has been reported so far for V ».

In fact, Kenik and Mitchell [3], had determined the TDE for vanadium in various directions by studying the onset of loop formation under irradiation in an electron microscope, after the investigation of poly-crystalline V by Miller and Chaplin [4] through resistivity measurements. The former had found a minimum TDE of 30 eV in the <100> direction, slightly higher than the polycrystal value of 26 eV of ref.[4] (confirmed later by Jung and Lucki [5] on V(300ppm Zr)) and far above the 13 eV computed by the authors of ref.[1]. This strong discrepancy ought certainly have been discussed by them, in view of possible methodological or other problems (bad potential?).

Instead, they do compare their results with « available » experimental data, but in a rather curious way. Fig.2 of ref.[1] includes, in addition to the set of their calculated TDE in various directions, also experimental data for Fe (ref. [Z-22][6] and for Mo (ref. [Z-23][7]. Now, ref. [Z-22][6] describes the experimental determination of the TDE in single crystal *tantalum*, where the authors of [6] compare their data with published *simulation* results by Erginsoy et al. [8] on *iron*: this then leads to the *experimental* points on Fe in Fig.2 of ref.[1].

Similarly, it should be tempting to compare the computed Frenkel pair separation distance of 2.6 $a_0$ to 8.1 $a_0$ (end of page 3 in [1]) with the spontaneous recombination volume experimentally determined in vanadium to 690-810 atomic volumes by Vajda and Biget [9] and related, together with other bcc metals, to their compressibility [10]; both articles unknown to the authors of [1].

In the same context, it is somewhat puzzling that the authors' calculations yield a practically temperature independent TDE, though they find it «not surprising» (last paragraph on p. 4 of [1]) in view of the «very much greater magnitude of the TDE than thermal energies». Now, it had been found experimentally, in a series of metals, that the TDE *de*creased substantially with increasing temperature. Thus, Urban and Yoshida [11] observed a TDE in Cu decreasing from 17 eV at 70 K (as compared to 19 eV at 4.2 K) to 10.5 eV at 550 K, while Zag and Urban [12] have found a TDE in Mo decreasing continuously from 34.5 eV at 70 K to 26.5 eV at 550 K. This temperature dependence was explained convincingly through an increasing escape probability from the spontaneous recombination volume at higher T (via interaction with thermal phonons) and hence increasing stability of the produced Frenkel pairs. It is, therefore, troubling to note the quasi-absence of such an effect in the present computer calculations up to 900 K. (This thermal stability does not concern the recovery of Frenkel pairs in annealing studies and, therefore, involves only those having escaped close-pair recombination.)

Concluding, it is surprising that workers involved in damage studies in *fusion* reactor materials should not be interested in the rich literature available on *fission* reactor materials, which had been investigated extensively 30 years ago. In particular, the TDE determination in metals has been treated in great detail in a review article by myself in 1977 [13] and summarized 15 years later in a Landolt-Börnstein volume [14]; this one quoted by the authors of [1] as ref. [Z-25] where they are only interested by the low migration energy of the self-interstitial atom on page 173 but not by the TDE determinations given on page 16!

References:

[1] L.A. Zepeda-Ruiz, S. Han, D.J. Srolovitz, R. Car, B.D. Wirth, Phys. Rev. B **67**, 134114 (2003).

[2] S.J. Zinkle, C. Kinoshita, J. Nucl. Mat. **251**, 200 (1997).

[3] E.A. Kenik, T.E. Mitchell, Phil. Mag. **32**, 815 (1975).

[4] M.G. Miller, R.L. Chaplin, Rad. Eff. **22**, 107 (1974).

[5] P. Jung, G. Lucki, Rad. Eff. **26**, 99 (1975).

[6] P. Jung, W. Schilling, Phys. Rev. B **5**, 2046 (1972).

[7] F. Maury, P. Vajda, M. Biget, A. Lucasson, P. Lucasson, Rad. Eff. **25**, 175 (1975).

[8] C. Erginsoy, G.H. Vineyard, A. Englert, Phys. Rev. **133**, A595 (1964).

[9] P. Vajda, M. Biget, Phys. Status Solidi (a) **23**, 251 (1974).

[10] M. Biget, R. Rizk, P. Vajda, A. Bessis, Solid State Comm. **16**, 949 (1975).

[11] K. Urban, N. Yoshida, Phil. Mag. A **44**, 1193 (1981).

[12] W. Zag, K. Urban, Phys. Status Solidi(a) **76**, 285 (1983).

[13] P. Vajda, Rev. Mod. Phys. **49**, 481 (1977).

[14] P. Ehrhart et al. in *Atomic Defects in Metals*, (ed. by H. Ullmaier), Landolt-Börnstein, New Series Group III, vol. 25 (Springer, 1991).